\documentclass[prd,aps,showpacs,twocolumn,longbibliography]{revtex4-1}
\usepackage{bm}
\usepackage{amsmath}
\usepackage{amssymb}
\usepackage{graphicx}
\usepackage{slashed}
\usepackage{accents}
\usepackage{mathrsfs}
\usepackage{braket}
\usepackage{units}
\usepackage{upgreek}

\newcommand{\highlight}[1]{\textit{#1}}
\renewcommand{\highlight}[1]{#1}

\usepackage[breaklinks=true,colorlinks=true,linkcolor=blue, citecolor=blue, urlcolor=blue]{hyperref}
\usepackage{breakurl}

\DeclareSymbolFont{eulerscript}{U}{eur}{m}{n}
\DeclareSymbolFontAlphabet{\matheuler}{eulerscript}
\DeclareMathAlphabet{\boldgreek}{OML}{zplm}{b}{it}
\newcommand{\I}{i}
\newcommand{\nfrac}[2]{{#1}/{#2}}
\newcommand{\one}{\mathbf{1}}
\newcommand{\eps}{\epsilon}
\newcommand{\mc}[1]{\mathcal{#1}}
\newcommand{\spvec}[1]{\boldsymbol{#1}}
\newcommand{\lb}{\left(}
\newcommand{\rb}{\right)}
\newcommand{\rsb}{\right]}
\newcommand{\lsb}{\left[}
\newcommand{\rcb}{\right\}}
\newcommand{\lcb}{\left\{}
\newcommand{\abs}[1]{\left|#1\right|}
\DeclareMathOperator{\tr}{\mathbf{tr}}
\DeclareMathOperator{\diag}{diag}
\DeclareMathOperator{\Ai}{Ai}
\newcommand{\lplus}{{{}+}}
\newcommand{\lminus}{{{}-}}
\newcommand{\lone}{{\scalebox{.64}{$\matheuler{I}$}}}
\newcommand{\ltwo}{{\scalebox{.64}{$\matheuler{II}$}}}
\newcommand{\lperp}{\perp}
\usepackage{slashed}\newcommand{\s}[1]{\slashed{#1}}
\newcommand{\Ftilde}{\mathfrak{F}}

\newcommand{\sdist}{\kern 0.20em}
\renewcommand{\eqref}[1]{Eq.\sdist(\ref{#1})}
\newcommand{\figref}[1]{Fig.\sdist\ref{#1}}
\newcommand{\appref}[1]{App.\sdist\ref{#1}}
\newcommand{\secref}[1]{Sec.\sdist\ref{#1}}

\begin{document}

\title{Semiclassical picture for electron-positron photoproduction in strong laser fields}

\author{Sebastian \surname{Meuren}}
\email{s.meuren@mpi-hd.mpg.de}

\author{Christoph H. \surname{Keitel}}
\email{keitel@mpi-hd.mpg.de}

\author{Antonino \surname{Di Piazza}}
\email{dipiazza@mpi-hd.mpg.de}

\affiliation{Max-Planck-Institut f\"ur Kernphysik, Saupfercheckweg 1, D-69117 Heidelberg, Germany}
\date{\today}

\begin{abstract}
The nonlinear Breit-Wheeler process is studied in the presence of strong and short laser pulses. We show that for a relativistically intense plane-wave laser field many features of the momentum distribution of the produced electron-positron pair like its extension, region of highest probability and carrier-envelope phase effects can be explained from the classical evolution of the created particles in the background field. To this end an intuitive semiclassical picture based on the local constant-crossed field approximation applied on the probability-amplitude level is established \highlight{and compared with the standard approach used in QED-PIC codes}. \highlight{The main difference is the substructure of the spectrum, which results from interference effects between macroscopically separated formation regions}. In order to compare the predictions of the semiclassical approach with exact calculations, a very fast numerical scheme is introduced. It renders the calculation of the fully differential spectrum on a grid which resolves all interference fringes feasible. Finally, the difference between classical and quantum absorption of laser four-momentum in the process is pointed out and the dominance of the former is proven. \highlight{As a self-consistent treatment of the quantum absorption is not feasible within existing QED-PIC approaches, our results provide reliable error estimates relevant for regimes where the laser depletion due to a developing QED cascade becomes significant}.    
\end{abstract}

\maketitle

\section{Introduction}
\label{sec:introduction}

From a conceptual point of view the transformation of light into matter is one of the most appealing physical processes. The possibility to create an electron-positron pair by merging two real photons (Breit-Wheeler process \cite{breit_collision_1934}) is a direct manifestation of the equivalence of mass and energy, postulated first by Einstein. Hitherto, however, this process has not been observed in a laboratory. Experimentally, electron-positron photoproduction was studied only indirectly via the trident process, i.e.\ by colliding a highly relativistic electron beam with an optical laser in the SLAC E-144 experiment \cite{burke_positron_1997} \highlight{(the experimental findings could be explained within the two-step approximation, i.e. by the assumption that the electrons first radiate real gamma photons which subsequently decay into pairs via the Breit-Wheeler process). To understand the results of the measurement theoretically, nonlinear effects (i.e. the simultaneous absorption of several laser photons) must be taken into account \cite{burke_positron_1997,hu_complete_2010,ilderton_trident_2011,king_trident_2013}.} 
\begin{figure}[b!]
\centering
\includegraphics{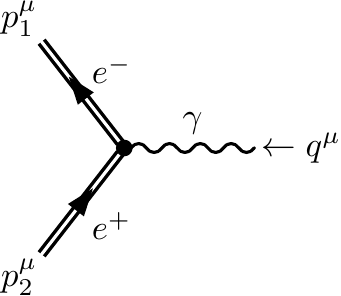}
\caption{\label{fig:pcdiagram}Leading-order Feynman diagram for electron-positron photoproduction inside a plane-wave background field (nonlinear Breit-Wheeler process). The double lines represent Volkov states (solutions of the interacting Dirac equation, which take the plane-wave background field into account exactly \cite{landau_quantum_1981,volkov_ueber_1935}), the wiggly line the incoming photon. As long as the total pair-production probability is much smaller than unity, the spectrum for the final particles is determined to a good accuracy by simply evaluating this diagram and neglecting radiative corrections \cite{meuren_polarization-operator_2015}.}
\end{figure}

Due to the continuous improvement of laser technology the experimental observation of the nonlinear generalization of the Breit-Wheeler process (see \figref{fig:pcdiagram}) is now within reach, e.g., at upcoming high-power laser facilities like the APOLLON-10P laser \cite{cheriaux_apollon_2012}, the Extreme Light Infrastructure~\cite{ELI}, the Vulcan $\unit[10]{PW}$ laser \cite{andrey_lyachev_10pw_2011} at the Central Laser Facility~\cite{CLF}, and the Exawatt Center for Extreme Light Studies~\cite{XCELS}. Therefore, this process has recently been considered by several authors \cite{wollert_tunneling_2014,krajewska_breit-wheeler_2014,titov_breit-wheeler_2013,fedotov_pair_2013,jansen_strongly_2013,king_photon_2013,nousch_pair_2012,titov_enhanced_2012,krajewska_breit-wheeler_2012,hebenstreit_pair_2011,ipp_streaking_2011,orthaber_momentum_2011,tuchin_non-linear_2010,heinzl_finite_2010,bulanov_multiple_2010,dunne_catalysis_2009,schutzhold_dynamically_2008} (see also the reviews \cite{battesti_magnetic_2013,di_piazza_extremely_2012,ruffini_electronpositron_2010,ehlotzky_fundamental_2009,mourou_optics_2006,marklund_nonlinear_2006,ritus_1985,mitter_quantum_1975}). 

The decay of a photon into an electron-positron pair is an intrinsic quantum process, which has no classical analogue and must be described in the realm of quantum field theory, e.g., by calculating the corresponding $S$-matrix element. This implies that we can only determine the total probability for the decay and the asymptotic momentum distribution for the final particles.

However, it is well known that in the case of a plane-wave laser field with electric field amplitude $E_0$ and central angular frequency $\omega$ the formation region of the basic QED processes nonlinear Compton scattering and nonlinear Breit-Wheeler pair production are $\xi$-times smaller than the laser period in the ultra-relativistic regime $\xi \gg 1$. Here, $\xi = \nfrac{|e|E_0}{(m\omega c)}$ is the classical intensity parameter, with $e<0$ and $m$ denoting the electron charge and mass, respectively \cite{di_piazza_extremely_2012,ritus_1985}. Hence, the total probability for nonlinear Compton scattering and nonlinear Breit-Wheeler pair production can be calculated by applying the local constant-crossed field approximation, i.e.\ by averaging the corresponding probability in a constant-crossed field over the laser pulse \cite{meuren_polarization-operator_2015,di_piazza_extremely_2012,ritus_1985,nikishov_pair_1967,narozhny_quantum_1965,nikishov_quantum_1964,nikishov_quantum2_1964,reiss_absorption_1962}. 

As pointed out by Ritus \cite{ritus_1985}, this procedure is justified for the calculation of total probabilities but fails in general for the momentum distribution of the final particles (this has also been recently observed numerically in \cite{harvey_testing_2015} for nonlinear Compton scattering). The reason are interference effects arising from processes occurring at macroscopically separated space-time points, which are neglected from the beginning when the averaging procedure is applied to the probabilities. As a laser field is oscillatory, each cycle typically contains two formation regions where the electron-positron pair is created with the same asymptotic quantum numbers. An interference between different pathways is therefore expected from general principles similar to a multi-slit experiment. 

\highlight{The fact that the substructure of the spectrum can be attributed to interference effects is well known from other processes. For nonlinear Thomson (Compton) scattering this was, e.g., reported in Refs. \cite{seipt_nonlinear_2011,mackenroth_nonlinear_2011,heinzl_beam-shape_2010,krafft_spectral_2004,narozhnyi_photon_1996}, for Schwinger pair production interference effects are, e.g., discussed in Refs. \cite{akkermans_ramsey_2012,dumlu_interference_2011,dumlu_stokes_2010,hebenstreit_momentum_2009}.}

Nonetheless, the average over the laser pulse shape on the probability level \highlight{(incoherent summation of all possible processes)} rather than the amplitude level  \highlight{(coherent summation of all possible processes)} is the state-of-the-art approach for the implementation of strong-field QED processes in so-called particle-in-cell (PIC) codes \cite{edwards_strongly_2016,grismayer_seeded_2015,tamburini_laser-pulse-shape_2015,vranic_particle_2015,gonoskov_extended_2015,gelfer_optimized_2015,green_simla_2015,lobet_ultrafast_2015,ridgers_modelling_2014,tang_qed_2014,bashmakov_effect_2014,mironov_collapse_2014,narozhny_creation_2014,brady_synchrotron_2014,ji_energy_2014,bulanov_electromagnetic_2013,lobet_modeling_2013,ridgers_dense_2013,ridgers_dense_2012,nerush_kinetic_2011,sokolov_numerical_2011,duclous_monte_2011,elkina_qed_2011,nerush_laser_2011,fedotov_limitations_2010,bulanov_schwinger_2010,kirk_pair_2009,bell_possibility_2008}. Therefore, it is desirable to revisit this approach and show how it could be extended if a higher precision becomes necessary. To this end we study here for the first time the quantitative influence of interference effects on the spectra for electron-positron photoproduction in the semiclassical regime.

By applying a stationary-phase analysis to the leading-order $S$-matrix element for electron-positron photoproduction it is shown below that in the strong-field regime $\xi\gg 1$ all qualitative features of the spectrum (including the substructure caused by interference effects) can be understood using the following semiclassical description: at each laser phase the pair-production probability \textit{amplitude} is calculated by employing the local constant-crossed field approximation. It predicts the local momentum spectrum of the pair immediately after the particles are brought on shell. The latter is then employed as initial condition for the classical propagation which provides the asymptotic momentum distribution. Finally, the probability for the production of a pair with given asymptotic momenta is obtained by squaring the probability amplitude, taking the interference between pairs which have the same asymptotic momenta but originating from different formation regions into account. 

This intuitive picture provides a clear physical reason for many properties of the asymptotic momentum distribution like its extension and the regions of highest probability. Moreover, it explains the strong dependence of the spectrum on the carrier-envelope phase (CEP) for ultra-short laser pulses reported in \cite{krajewska_breit-wheeler_2012}. In the context of vacuum pair production classical features have been identified in \cite{chott_classical_2007}. Furthermore, a similar semiclassical analysis has been carried out in \cite{mackenroth_nonlinear_2013,mackenroth_nonlinear_2011,mackenroth_determining_2010} to explain various features of the emission spectra for nonlinear Compton scattering and has been exploited in \cite{mackenroth_determining_2010} to put forward a scheme for determining the CEP in ultra-short and ultra-intense laser beams via nonlinear single-Compton scattering. 

We point out that electron-positron photoproduction has a lot of commonalities with laser-induced ionization processes. In fact, the procedure outlined above is closely related to similar approaches used in atomic physics to describe the time evolution of an electron after tunnel ionization \cite{kohler_frontiers_2012,corkum_plasma_1993}. 

Interestingly, this semiclassical three-step model allows us to distinguish between a ``classical'' and a ``quantum'' absorption of laser four-momentum in the process. As the decay of a photon into an electron-positron pair is forbidden in vacuum, a certain amount of laser four-momentum must be absorbed initially to bring the massive particles on shell (this part will be called quantum absorption here, see \secref{sec:classicalvsquantumabsorption}). Afterwards, the charged particles are further accelerated by the laser field, which implies a classical energy-momentum transfer. Below, these two processes are distinguished for the first time and it is shown that classical absorption dominates for $\xi\gg 1$. Correspondingly, the laser is predominantly depleted from the classical energy-momentum transfer.

To establish the validity of the outlined semiclassical approach, we compare its predictions with a full numerical calculation of the leading-order $S$-matrix element (see \figref{fig:pcdiagram}). In this way we show that already for $\xi \gtrsim 5$ the interference substructure obtained from the local constant-crossed field approximation applied on the probability amplitude level is in very good agreement with the full numerical result. 

As the $S$-matrix integrals are highly oscillating in the regime $\xi\gg1$, the numerical calculation is, in principle, a challenging task \cite{titov_enhanced_2012,krajewska_breit-wheeler_2012}. We present here a new scheme, which is substantially more efficient than other employed methods \highlight{(see \secref{sec:ftmasterintegrals})}. Hence, it becomes feasible to evaluate the three-dimensional differential probability on a grid which is fine enough to completely resolve the interference substructure of the spectrum. To obtain the total pair-production probability, however, the numerical evaluation of the compact expressions derived in \cite{meuren_polarization-operator_2015} is still much faster.

Note that for nonlinear Compton scattering the final phase space has been studied in \cite{seipt_asymmetries_2013,boca_electron_2012,krajewska_compton_2012,boca_thomson_2010} for moderate values of the parameter $\xi$ ($\xi\lesssim 10$) and the computation of total probabilities can also be significantly simplified by employing the method proposed in \cite{dinu_exact_2013}.

\highlight{The paper is organized as follows. In \secref{sec:ppp} our notation is introduced and the well-known leading-order expression for the pair-production probability [see \eqref{eqn:paircreation_probabilitycanonincalparametrization}] is expressed using Lorentz-invariant momentum parameters (see \secref{sec:invariantmomentumparameters} for details). Furthermore, the numerical scheme for the numerical evaluation of the pair-production probability is described in \secref{sec:ftmasterintegrals}. Afterward, the main new result of the present paper, the disentanglement between classical and quantum aspects of the pair-production process, is presented in \secref{sec:semiclassicalpicture}.}

From now on we use natural units $\hbar = c = 1$ and Heaviside-Lorentz units for charge [$\alpha = \nfrac{e^2}{(4\pi)} \approx \nfrac{1}{137}$ denotes the fine-structure constant], the notation agrees with \cite{meuren_polarization_2013,meuren_polarization-operator_2015}.

\section{Pair production probability}
\label{sec:ppp}

In this paper we consider the decay of a photon with four-momentum $q^\mu$ ($q^2=0$) and polarization four-vector $\eps^\mu$ into an electron and a positron with four-momentum $p_1^\mu$ and $p_2^\mu$, respectively ($p_i^2=m^2$). In vacuum, this transition is forbidden by energy-momentum conservation (for all photon energies). Inside an electromagnetic background field, however, the so-called (nonlinear) Breit-Wheeler process is in general allowed (see \figref{fig:pcdiagram}). Here, we focus on plane-wave laser fields described by the classical field tensor
\begin{gather}
\label{eqn:fieldtensor}
F^{\mu\nu}(\phi) = \sum_{i=1,2} f_i^{\mu\nu} \psi_i'(\phi),
\quad
f_i^{\mu\nu} = k^\mu a_i^\nu - k^\nu a_i^\mu,
\end{gather}
where $\phi=kx$ is the laser phase, $x^\mu$ the position four-vector, $k^\mu$ the characteristic four-momentum of the laser photons and $a_i^\mu$ and $\abs{\psi_i'(\phi)} \lesssim 1$ characterize the strength and the shape of the field, respectively, along the two possible polarization directions [$i=1,2$; $\psi_i(\pm\infty)=\psi'_i(\pm \infty)=0$; the prime denotes the derivative of a function with respect to the argument]. Furthermore, we introduce the classical intensity parameters 
\begin{gather}
\xi_i = \frac{\abs{e}}{m} \sqrt{-a_i^2},
\quad
\xi = \sqrt{\xi_1^2 + \xi_2^2}
\end{gather}
and the quantum-nonlinearity parameter $\chi = (\nfrac{kq}{m^2}) \xi$ (note that $kq \geq 0$ as $q^2=0$ here). 

If a laser field is sufficiently intense and depletion effects are negligible, it can be considered as a classical background field. Correspondingly, the pair-production probability is obtainable using the formalism of strong-field QED in the Furry picture \cite{furry51} (for more details see, e.g., \cite{meuren_nonlinear_2015,battesti_magnetic_2013,di_piazza_extremely_2012,ruffini_electronpositron_2010,ehlotzky_fundamental_2009,mourou_optics_2006,marklund_nonlinear_2006,dittrich_probingquantum_2000,fradkin_quantum_1991,ritus_1985,mitter_quantum_1975} \highlight{and the literature cited in the introduction}).

\subsection{Invariant momentum parameters}
\label{sec:invariantmomentumparameters}

Using the canonical light-cone basis (see \appref{sec:lccappendix} for more details), we introduce the Lorentz-invariant momentum parameters $r$, $t_1$ and $t_2$
\begin{subequations}
\label{eqn:momentumincanonicallcb}
\begin{align}
p^\mu_1  &=  r' q^\mu + s' k^\mu + t'_1  m \Lambda_1^\mu + t'_2  m \Lambda_2^\mu,
\\
p^\mu_2  &= -r q^\mu - s k^\mu - t_1 m \Lambda_1^\mu - t_2 m \Lambda_2^\mu.
\end{align}
\end{subequations}
Here, $t_i'=t_i$ and $r'=r+1$ due to momentum conservation and the quantities
\begin{gather}
\label{eqn:momentumincanonicallcbssprime}
s = \frac{1}{2r} \frac{m^2}{kq} (1 + t_1^2 + t_2^2),
\quad
s' = \frac{1}{2r'} \frac{m^2}{kq} (1 + t_1^2 + t_2^2)
\end{gather}
are determined by the on-shell conditions $p_1^2=p_2^2=m^2$. We point out that we consider only plane-wave fields with a finite duration. Correspondingly, the four-momentum $q^\mu$ ($p_{1}^\mu$, $p_{2}^\mu$) denotes the asymptotic four-momentum in vacuum before (after) the interaction with the laser pulse. In particular, we do not introduce dressed masses and momenta for the electron and the positron \cite{di_piazza_extremely_2012,ritus_1985}.

For later convenience we also define the quantities \cite{ritus_1985}
\begin{subequations}
\label{eqn:momentumincanonicallcc_Rw}
\begin{align}
R &= r+\frac{1}{2} = \frac{kp_1-kp_2}{2kq} = \frac{kp_1-kp_2}{2(kp_1+kp_2)},
\\
w &= -\frac{1}{r(r+1)} = \frac{4}{1-4R^2} = \frac{(kq)^2}{(kp_1) (kp_2)}
\end{align}
\end{subequations}
to characterize how the initial photon four-momentum is split between the outgoing particles (unlike $r$ and $r'$ the parameters $R$ and $w$  are anti-symmetric and symmetric with respect to the electron and the positron four-momentum, respectively). As $kq > 0$, $kp_i>0$ and $r'=r+1$ we conclude from \eqref{eqn:momentumincanonicallcb} that $r\in (-1,0)$, $r'\in (0,1)$, $R \in (-\nfrac{1}{2},+\nfrac{1}{2})$ and $w \in [4,\infty)$.

Note that the map from $R\to w$ has no inverse (i.e.\ the information about the sign of $R$ is lost). Correspondingly, the quantities $kq$ and $w$ [see \eqref{eqn:momentumincanonicallcc_Rw}] specify $kp_1$ and $kp_2$ uniquely up to the sign of $R$. However, we will later see that this sign does not influence the spin-summed pair-production probability.

\highlight{Using the Lorentz-invariant momentum parameters $r$ ($R$), $t_1$ and $t_2$ to describe the asymptotic momenta of the created electron-positron pair has the advantage that they characterize the process in a frame-independent way. In particular, we do not have to work in a frame where the collision is head on. However, in this frame the parameters $t_i$ have a simple interpretation, as they measure the transverse momentum of the pair in units of the electron rest mass, whereas $R$ measures in general how the initial photon four-momentum $q^\mu$ is distributed between the electron and the positron.} 

Note that in the seminal papers \cite{nikishov_pair_1967,ritus_1985} the differential pair-production rate inside a constant-crossed field is expressed with respect to the two parameters $u$ and $\tau$, which are related to the parameters $w$ and $t_2$ introduced here as follows
\begin{subequations}
\begin{gather}
w
=
\frac{qf^2q}{\sqrt{(p_1f^2p_1)(p_2f^2p_2)}}
= 
4u,
\\
t_2^2 = \tau^2,
\quad
\tau = \frac{p_1 f^* p_2}{m \sqrt{qf^2q}}
\end{gather}
\end{subequations}
($f^{\mu\nu} = f_1^{\mu\nu}$). Therefore, the comparison between the results obtained here for a laser pulse and those reported previously for a constant-crossed field is straightforward in canonical light-cone coordinates.

The reason why the parameter $t_1$ does not appear in the final expression for the pair-production rate in a constant-crossed field is related to the fact that $t_1$ is not a constant of motion (with respect to the classical equations of motion, see \secref{sec:classicalinterpretationofthestatpoints}). Correspondingly, the final (asymptotic) value of $t_1$ will depend on the subsequent classical evolution of the charged particles and therefore on the nonlocal properties of the background field (see also the discussion in \secref{sec:initialconditionsclassicalpropagation}). To avoid this problem, $t_1$ is not specified for a constant-crossed background field and only pair-production rates (not total probabilities) are calculated. For a laser pulse with finite duration, however, one obtains a total probability rather than a rate. Furthermore, as the particles leave the background field at a certain point, it is possible to specify the momentum distribution of the final particles with respect to the parameter $t_1$.

Finally, we note that the amount $n k^\mu$ of absorbed laser four-momentum is related to the introduced Lorentz invariant momentum parameters as follows
\begin{gather}
\label{eqn:momentumconservationpaircreation}
p_1^\mu + p_2^\mu = q^\mu + n k^\mu,
\quad
n = \frac{1}{2} w \frac{m^2}{kq} (1 + t_1^2 + t_2^2).
\end{gather}
We will later show that this (asymptotic) value contains both a classical and a quantum contribution (see \secref{sec:classicalvsquantumabsorption}). As the laser-photon energy is only well defined for monochromatic fields, $n$ is in general not an integer and only $n k^\mu$ is a meaningful quantity. Nevertheless, we will call $n$ the number of absorbed laser photons.

\subsection{Total and differential probability}
\label{sec:totalanddifferentialprobability}

Using the parameters introduced in \secref{sec:invariantmomentumparameters}, the total probability $W(q,\eps)$ for the decay of a gamma photon with four-momentum $q^\mu$ ($q^2=0$) and polarization four-vector $\eps^\mu$ ($\eps{}q =0$) into an electron-positron pair (see \figref{fig:pcdiagram}) can be written as (see, e.g., \cite{meuren_polarization-operator_2015} \highlight{and the literature cited in the introduction})
\begin{subequations}
\label{eqn:paircreation_probabilitycanonincalparametrization}
\begin{align}
W(q,\eps)
&=
\int_{-\nfrac{1}{2}}^{+\nfrac{1}{2}} dR \int^{+\infty}_{-\infty} dt_1 dt_2 \, \frac{d^3W}{dR dt_1 dt_2},
\\
\frac{d^3W}{dR dt_1 dt_2}
&=
\frac{m^2}{(kq)^2} \frac{w}{8} \, \frac{1}{(2\pi)^3} \sum_{\text{spin}} \abs{\mc{M}(p_1,p_2;q)}^2,
\end{align}
\end{subequations}
where $\I\mc{M}(p_1,p_2;q) = \eps_\mu \, \bar{u}_{p_1} \mc{G}^\mu(p_1,q,-p_2) v_{p_2}$ is the reduced $S$-matrix element for the process %
\footnote{Note that in Eq.\,(3) and Eq.\,(A19) of \cite{meuren_polarization-operator_2015} the $\I$ is erroneously missing.}
and we sum with respect to the final spin quantum numbers of the created particles (for simplicity they are not indicated). Here, $u_{p_1}$ and $v_{p_2}$ denote the Dirac spinors for the electron and the positron, respectively, and $\mc{G}^\mu$ the nonsingular part of the laser-dressed vertex [see \eqref{eqn:dressedvertexfinal_nonsingularpartcanonicalparametrization}]. 

In \eqref{eqn:paircreation_probabilitycanonincalparametrization} the phase-space integrals are written in terms of the invariant momentum parameters defined in Eqs.\sdist(\ref{eqn:momentumincanonicallcb}) and (\ref{eqn:momentumincanonicallcc_Rw}) and $\nfrac{d^3W}{(dR dt_1 dt_2)}$ represents the differential pair-creation probability with respect to those parameters. In order to calculate it we note that 
\begin{multline}
\label{eqn:paircreation_spinsummedmatrixelement}
\sum_{\text{spin}} \abs{\mc{M}(p_1,p_2;q)}^2
=
\eps_\mu \eps^*_\nu \tr \mc{G}^\mu(p_1,q,-p_2) \\\times (\s{p}_2-m) \bar{\mc{G}}^\nu(p_1,q,-p_2) (\s{p}_1+m).
\end{multline}
For on-shell momenta the nonsingular part of the dressed vertex is given by (see \cite{meuren_polarization_2013,meuren_polarization-operator_2015} for more details)
\begin{multline}
\label{eqn:dressedvertexfinal_nonsingularpartcanonicalparametrization}
\mc{G}^\rho(p_1,q,-p_2)
=
(-\I e) \bigg\{
\gamma_\mu  \Big[\mathfrak{G}_{0} g^{\mu\rho}
+ 
\sum_{j=1,2} \, (G_1 \mathfrak{G}_{j,1} f^{\mu\rho}_j
\\+ 
G_2 \mathfrak{G}_{j,2} f_j^{2\mu\rho}) \Big]
+ 
\I \gamma_\mu \gamma^5 \sum_{j=1,2} \, G_3 \mathfrak{G}_{j,1} f_j^{*\mu\rho} \bigg\},
\end{multline}
where
\begin{gather}
\label{eqn:dressedvertex_Givswkq}
\begin{gathered}
G_1(R,kq) = e \frac{Rw}{kq},
\quad
G_2(R,kq) = -\frac{e^2}{2} \, \frac{w}{(kq)^2},
\\
G_3(R,kq) = -\frac{e}{2} \, \frac{w}{kq}.
\end{gathered}
\end{gather}
Furthermore, the so-called master integrals $\mathfrak{G}_{0} = \mathfrak{G}_{0}(w,t_1,t_2)$ and $\mathfrak{G}_{j,l} = \mathfrak{G}_{j,l}(w,t_1,t_2)$ are given by (the notation agrees with \cite{meuren_nonlinear_2015})
\begin{subequations}
\label{eqn:dressedvertexmasterintegrals}
\begin{align}
\mathfrak{G}_{0}
&=
\int_{-\infty}^{+\infty} d\phi \, e^{\I \widetilde{S}_\Gamma(w,t_1,t_2;\phi)},
\\
\mathfrak{G}_{j,l}
&=
\int_{-\infty}^{+\infty} d\phi \, [\psi_j(\phi)]^l  e^{\I \widetilde{S}_\Gamma(w,t_1,t_2;\phi)},
\end{align}
\end{subequations}
with the nonlinear, field-dependent phase 
\begin{subequations}
\label{eqn:dressedvertex_reducedphase_onshell}
\begin{gather}
\widetilde{S}_\Gamma(w,t_1,t_2;\phi) = \frac{w}{2} \frac{m^2}{kq} \mathfrak{S}_\Gamma(t_1,t_2;\phi),
\end{gather}
\begin{multline}
\mathfrak{S}_\Gamma(t_1,t_2;\phi) = (1 + t_1^2 + t_2^2) \phi 
\\+  \sum_{i=1,2} \, \int_{-\infty}^{\phi} d\phi'\, \Big[ \xi^2_i \psi^2_i(\phi') -  2 t_i \xi_i \psi_i(\phi') \Big].
\end{multline}
\end{subequations}

From \eqref{eqn:dressedvertexfinal_nonsingularpartcanonicalparametrization} we conclude that the pair-creation probability can be calculated algebraically once the master integrals are known. As the integration range of $\mathfrak{G}_{j,l}(w,t_1,t_2)$ is naturally bounded, a numerical calculation is readily accomplished (for $l\neq0$, more details are provided in \secref{sec:ftmasterintegrals}). To determine $\mathfrak{G}_{0}(w,t_1,t_2)$ we integrate by parts and after neglecting the boundary terms we obtain the relation
\begin{multline}
\label{eqn:dressedvertexmasterintegrals_G0relation}
\mathfrak{G}_{0}(w,t_1,t_2)
=
- \frac{1}{2n} \frac{m^2}{kq} w \sum_{i=1,2}  \Big[\xi^2_i \mathfrak{G}_{i,2}(w,t_1,t_2) \\-  2t_i \xi_i \mathfrak{G}_{i,1}(w,t_1,t_2) \Big]
\end{multline}
[$n>0$, see \eqref{eqn:momentumconservationpaircreation}].

\subsection{Fourier-transformed master integrals}
\label{sec:ftmasterintegrals}

From \eqref{eqn:dressedvertex_reducedphase_onshell} we infer that the dependence of the master integrals $\mathfrak{G}_{j,l}(w,t_1,t_2)$ on $w$ is very simple. \highlight{As a consequence, their Fourier transforms $\tilde{\mathfrak{G}}_{j,l}(z,t_1,t_2)$ [defined in \eqref{eqn:dressedvertex_masterintegrals_wft}] can be calculated analytically.} To this end we have to consider $\mathfrak{G}_{j,l}(w,t_1,t_2)$ as a function of $w\in(-\infty,+\infty)$, even though only the parameter range $w\in[4,\infty)$ is important from a physical point of view [the master integrals are everywhere well defined, see \eqref{eqn:dressedvertexmasterintegrals}]. After interchanging the order of integration, we obtain the following representation
\begin{multline}
\label{eqn:dressedvertex_masterintegrals_wft}
\tilde{\mathfrak{G}}_{j,l}(z,t_1,t_2)
=
\int_{-\infty}^{+\infty} dw \, e^{-\I \frac{1}{2} w \frac{m^2}{kq} z} \, \mathfrak{G}_{j,l}(w,t_1,t_2)
\\= 
4\pi \, \frac{kq}{m^2} \frac{[\psi_j(\phi_{z})]^l}{\mathfrak{S}'_\Gamma(t_1,t_2;\phi_{z})},
\end{multline}
where the prime denotes the derivative with respect to the laser phase $\phi$ [see \eqref{eqn:dressedvertex_reducedphase_onshell}] and $\phi_{z}$ is the (unique) solution of the equation $\mathfrak{S}_\Gamma(w,t_1,t_2;\phi_{z}) = z$. The uniqueness of $\phi_{z}$ follows from the fact that
\begin{gather}
\label{eqn:dressedvertex_reducedphase_onshellderivative}
\mathfrak{S}'_\Gamma(t_1,t_2;\phi) = 1  +  \sum_{i=1,2} \, \big[ t_i - \xi_i \psi_i(\phi) \big]^2
\end{gather}
is always greater than zero on the real axis. Thus, the calculation of the Fourier-transformed master integrals $\tilde{\mathfrak{G}}_{j,l}(z,t_1,t_2)$ reduces to a root-finding problem (which is solvable numerically with low computational costs). 

\highlight{Once $\tilde{\mathfrak{G}}_{j,l}(z,t_1,t_2)$ is calculated on a grid in $z$ with sufficient resolution, the quantities $\mathfrak{G}_{j,l}(w,t_1,t_2)$, which are related to $\tilde{\mathfrak{G}}_{j,l}(z,t_1,t_2)$ by an inverse Fourier transform [see \eqref{eqn:dressedvertex_masterintegrals_wft}], can be calculated numerically on a grid in $w$ very efficiently by means of a \textit{single} Fast-Fourier Transform (FFT) \cite{press_numerical_2007,cooley_algorithm_1965}. Therefore, this approach reduces the problem of calculating $\mathfrak{G}_{j,l}(w,t_1,t_2)$ as a function of $w$, $t_1$ and $t_2$ on a three-dimensional grid to an effectively two-dimensional problem [from the viewpoint of computation costs, assuming that the root-finding problem in \eqref{eqn:dressedvertex_masterintegrals_wft} causes no significant overhead with respect to the FFT]. In comparison with a direct calculation of $\mathfrak{G}_{j,l}(w,t_1,t_2)$ [see \eqref{eqn:dressedvertexmasterintegrals}] using standard algorithms for highly, nonuniformly oscillating integrals, \eqref{eqn:dressedvertex_masterintegrals_wft} reduces the required computational effort substantially.}

Alternatively, one could also perform the change of variables $\phi\to z = \mathfrak{S}_\Gamma(w,t_1,t_2;\phi)$ in \eqref{eqn:dressedvertexmasterintegrals} and evaluate the master integrals directly via FFT. The change of variables is one-to-one as $\mathfrak{S}'_\Gamma(t_1,t_2;\phi) > 0$, see \eqref{eqn:dressedvertex_reducedphase_onshellderivative}. This approach has been applied in \cite{di_piazza_strong_2009} to the analogous problem of nonlinear Thomson scattering.

\section{Semiclassical picture}
\label{sec:semiclassicalpicture}

By combining all relations presented in \secref{sec:ppp}, the numerical evaluation of the leading-order $S$-matrix element for the nonlinear Breit-Wheeler process and the determination of the momentum distribution for the created particles is straightforward. However, we obtain no further physical insights into the pair-production process in this way, as the $S$-matrix does not reveal any details about the dynamics taking place inside the interaction zone. Therefore, an intuitive semiclassical picture is now developed, which is applicable for strong background fields ($\xi\gg 1$). Using optical lasers (photon energy $\omega\sim\unit[1]{eV}$) intensity parameters $\xi \gtrsim 100$ are accessible at existing and upcoming laser facilities \cite{yanovsky_ultra_2008,cheriaux_apollon_2012,ELI,andrey_lyachev_10pw_2011,CLF,XCELS}. In this regime the actual transformation from light to matter (which happens within a microscopically small formation region $\delta\phi \sim \nfrac{1}{\xi}$ in the laser phase) can be separated from the subsequent classical propagation of the created particles.

In order to verify the reliability of the semiclassical approach (which is to a large extend similar to the one used in PIC codes), we compare its predictions with full numerical calculations. To this end we consider a linearly polarized laser field [$\psi=\psi_1$, $\psi_2=0$, $\xi=\xi_1$] with the following pulse shape
\begin{gather}
\psi'(\phi) = \sin^2[\nfrac{\phi}{(2N)}] \, \sin(\phi + \phi_0)
\end{gather}
for $\phi \in [0,2\pi N]$ and zero otherwise. Here, $N$ denotes the number of cycles and $\phi_0$ the CEP of the pulse (the numerical values of the parameters used in the calculations are given in the captions of the figures). Furthermore, we assume that the incoming photon has parallel polarization ($\eps^\mu = \Lambda_1^\mu$, see \appref{sec:photonpolarizationappendix}). In this case the trace in \eqref{eqn:paircreation_spinsummedmatrixelement} is given by
\begin{multline}
\label{eqn:paircreation_linpoltracelambda1}
-\frac{1}{e^2} \Lambda_{1\mu}\Lambda_{1\nu} \tr[...]^{\mu\nu}
\\=
2m^2 (w-4) \big[ -\xi^2 \abs{\mathfrak{G}_{1,1}}^2 
+ 2\xi t_1 \Re (\mathfrak{G}_{0}^* \mathfrak{G}_{1,1}) \big]
\\+ 4m^2\abs{\mathfrak{G}_{0}}^2 \big[2t_1^2 
-(\nfrac{w}{2})(1+t_1^2+t_2^2) \big]
\end{multline}
and we denote the probability by $W_\parallel(q) = W(q,\eps=\Lambda_1)$ [see \eqref{eqn:paircreation_probabilitycanonincalparametrization}].

\subsection{Stationary-phase analysis}
\label{sec:stationaryphaseanalysis}

\highlight{To obtain an intuitive semiclassical picture, we apply now a stationary-phase analysis to the master integrals defined in \eqref{eqn:dressedvertexmasterintegrals}. However, our calculation does not precisely follow the method of steepest descent, which is the standard approach (see, e.g., \cite{ritus_1985,reiss_absorption_1962,narozhnyi_photon_1996} and also \cite{airy_intensity_1838}). Instead, it is shown that the integral  along the real line is dominated by those points where the second derivative of the phase vanishes. Of course, the final result agrees with the one obtained using the method of steepest descent, but the derivation is less complicated as it does not require the deformation of the integration contour within the complex plane.}

From \eqref{eqn:dressedvertex_reducedphase_onshell} we conclude that in the regime $\xi\gg 1$ the master integrals are in general highly oscillating. From the derivative of the phase 
\begin{gather}
\label{eqn:dressedvertex_reducedphase_onshellderivative_general}
\widetilde{S}'_\Gamma(w,t_1,t_2;\phi) = \frac{w}{2} \frac{m^2}{kq} \mathfrak{S}'_\Gamma(t_1,t_2;\phi)
\end{gather}
we infer that the master integrals have no stationary point on the real integration line (the prime denotes the derivative with respect to the laser phase $\phi$). Focusing on the case of linear polarization [see \eqref{eqn:dressedvertex_reducedphase_onshellderivative}]
\begin{gather}
\label{eqn:dressedvertex_reducedphase_onshellderivative_linpol}
\mathfrak{S}'_\Gamma(t_1,t_2;\phi) = 1  + t_2^2 + \big[ t_1 - \xi \psi(\phi) \big]^2,
\end{gather}
we find that the stationary points $\varphi_k^{\pm}$ of the phase $\widetilde{S}_\Gamma(\phi) = \widetilde{S}_\Gamma(w,t_1,t_2;\phi)$ [defined by $\widetilde{S}'_\Gamma(\varphi_k^{\pm})=0$] are complex and given by
\begin{gather}
\label{eqn:paircreation_statpointslinpol}
\psi(\varphi_k^{\pm}) = \frac{1}{\xi} \lb t_1 \pm \I \sqrt{1 + t_2^2} \rb.
\end{gather}
To obtain the leading-order approximation to the master integrals in the regime $\xi\gg 1$, one could apply the method of steepest descent, i.e.\ deform the integration contour inside the complex plane such that it passes through the stationary points (see, e.g., \cite{ritus_1985,reiss_absorption_1962,narozhnyi_photon_1996}). However, the desired result is derived much faster by noting that the stationary points $\varphi_k^{\pm}$ are located pairwise very close to the real line if $\abs{t_2} \lesssim 1$ \highlight{(analogously to nonlinear Compton scattering and other processes \cite{narozhnyi_photon_1996,mackenroth_nonlinear_2011,meuren_high-energy_2015} inside a plane-wave field;} for $\abs{t_2} \gg 1$ the pair-production probability is exponentially suppressed, see below). Mathematically, we have to deal with two stationary points which coalesce in the limit $\xi\to\infty$. For $\xi\gg 1$ the two stationary points nearly coalesce, a situation which is discussed, e.g., in \cite{chester_extension_1957} (see also \cite{meuren_nonlinear_2015}, App.\,H and \cite{olver_nist_2010}, Chap.\,36). 

\highlight{Due to the presence of two stationary points $\varphi_k^{\pm}$ close to each (quasi-) stationary point $\phi_k$ defined by 
\begin{gather}
\label{eqn:paircreation_statpointeq}
\psi(\phi_k) = \nfrac{t_1}{\xi},
\end{gather}
we expect that the integral along the real line is dominated by small formation regions $\delta\phi$ around the points $\phi_k$} [for a linearly polarized laser field; in general we obtain two equations $\psi_i(\phi_k) = \nfrac{t_i}{\xi_i}$ ($i=1,2$) which should be fulfilled simultaneously]. From \eqref{eqn:dressedvertex_reducedphase_onshellderivative_linpol} we conclude that the oscillation frequency of the phase is as small as possible at these points \highlight{(i.e. $\widetilde{S}_\Gamma''=0$)} because the dominating contribution to the oscillation frequency becomes stationary (an intuitive physical interpretation for this condition is given in \secref{sec:classicalinterpretationofthestatpoints}). In the following, we will call the points $\phi_k$ (and not $\varphi_k^{\pm}$) stationary points for simplicity (to stress the difference, the points $\varphi_k^{\pm}$ are called \textit{true} stationary points). Moreover, we assume that all stationary points $\phi_k$ are located sufficiently far away from each other, i.e.\ we ignore subtleties arising around the extremal points of $\psi(\phi)$ [note that pair production is ineffective in these regions, as $\psi'(\phi)$ is small].

As the main contribution to the master integrals arises from the regions around the phases $\phi_k$ where $t_1 \approx \xi \psi(\phi_k)$, we expand the phase $\widetilde{S}_\Gamma(\phi) = \widetilde{S}_\Gamma(w,t_1,t_2;\phi)$ [see \eqref{eqn:dressedvertex_reducedphase_onshell}] around a stationary point $\phi_k$ up to cubic order
\begin{gather}
\label{eqn:dressedvertex_linpol_phasesemiclassicapprox}
\begin{gathered}
\widetilde{S}_\Gamma(\phi) 
\approx
\widetilde{S}_\Gamma(\phi_k) + a (\phi-\phi_k) + \frac{1}{3} b (\phi-\phi_k)^3
\\
a = \frac{w}{2} \frac{m^2}{kq} (1 + t_2^2),
\quad
b = \frac{w}{2} \frac{m^2}{kq} [\xi \psi'(\phi_k)]^2.
\end{gathered}
\end{gather}
Here, we focus on the regime $\chi \gtrsim 1$ (where the pair-production probability is not exponentially suppressed, see below) and define the formation region $\delta\phi = \phi-\phi_k$ around $\phi_k$ by the requirement that the phase in \eqref{eqn:dressedvertex_linpol_phasesemiclassicapprox} remains of order one (see, e.g., \cite{baier_concept_2005,baier_electromagnetic_1994}). As $a\sim\xi$ and $b\sim \xi^3$, the formation region scales as $\delta\phi\sim\nfrac{1}{\xi}$ (we always assume $\xi\gg 1$ in this section). Correspondingly, both the linear and the cubic term have to be taken into account. Higher-order terms do not change the behavior of the phase significantly (within the formation region) and can be neglected to leading order. We will show in \secref{sec:classicalvsquantumabsorption} that the scaling $\delta\phi\sim\nfrac{1}{\xi}$ for the formation region is reasonable from a physical point of view, as the energy absorbed classically from the background field within $\delta\phi$ is sufficient to bring the particles on shell.

After the change of variables from $\phi$ to $t = \sqrt[3]{b} (\phi-\phi_k)$ the phase is approximately given by [see \eqref{eqn:dressedvertex_linpol_phasesemiclassicapprox}]
\begin{gather}
\widetilde{S}_\Gamma(\phi) 
\approx
\widetilde{S}_\Gamma(\phi_k) + x t + \frac{1}{3} t^3,
\end{gather}
where
\begin{gather}
\begin{gathered}
x = \frac{a}{\sqrt[3]{b}} = \lsb \frac{\nfrac{w}{2}}{\abs{\chi(\phi_k)}} \rsb^{\nfrac{2}{3}} (1+t_2^2),
\\
\frac{1}{\sqrt[3]{b}} =  \frac{2}{w} \frac{kq}{m^2} \lsb \frac{\nfrac{w}{2}}{\abs{\chi(\phi_k)}} \rsb^{\nfrac{2}{3}},
\end{gathered}
\end{gather}
and the absolute value of $\chi(\phi) = \chi \psi'(\phi)$ denotes the local value of the quantum-nonlinearity parameter $\chi = (\nfrac{kq}{m^2})\xi$ \cite{di_piazza_extremely_2012,ritus_1985}.

\begin{figure*}[t!]
\centering
\includegraphics{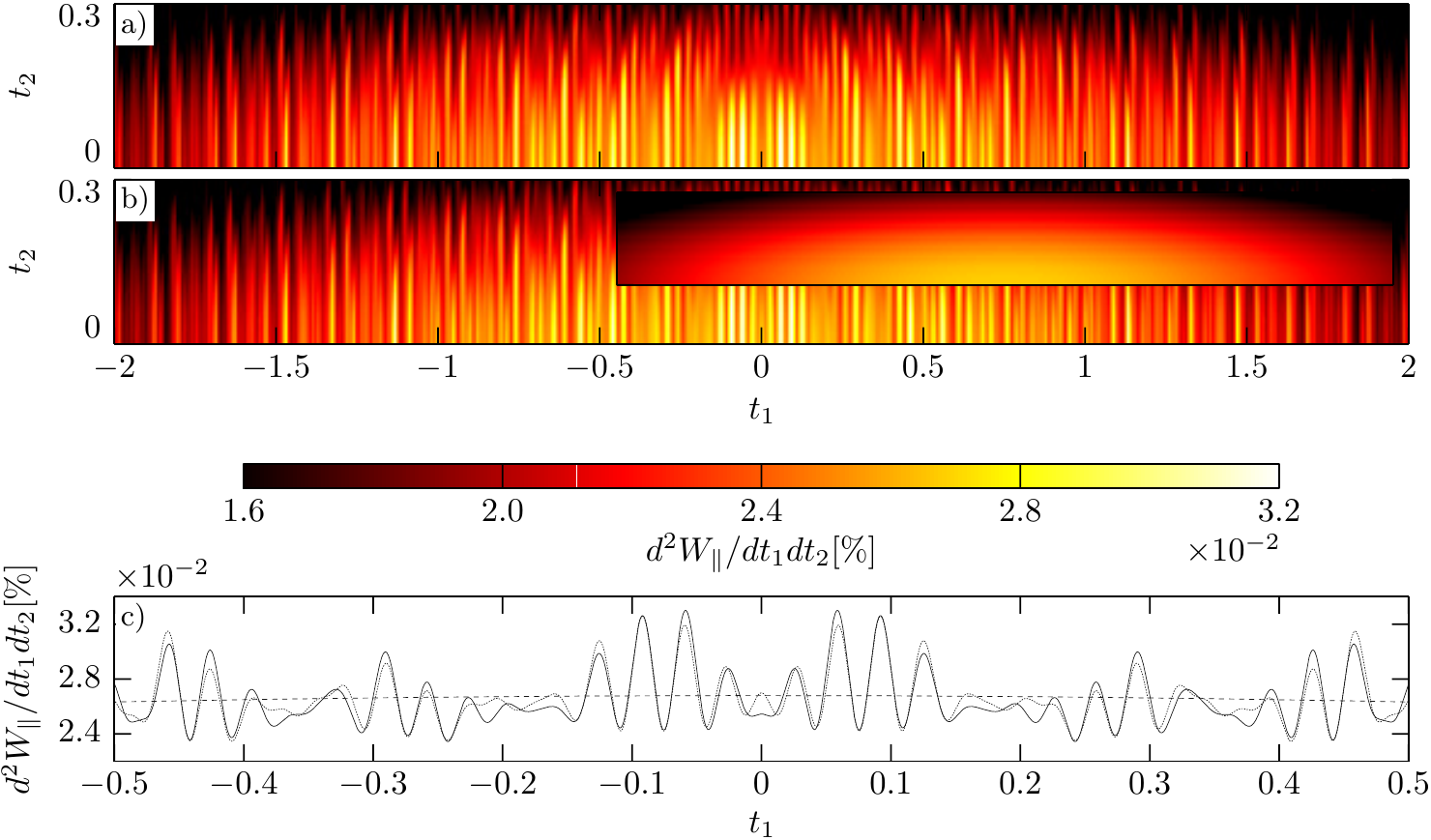}
\caption{\label{fig:spectrum_t1t2_N5interferencecompare}
Momentum distribution for the created electron-positron pair [see \eqref{eqn:paircreation_probabilitycanonincalparametrization}] for the parameters $\chi=1$, $\xi=5$, $N=5$ and $\phi_0=\nfrac{\pi}{2}$ [the longitudinal momentum characterized by $R$ is integrated numerically and the incoming photon has parallel polarization ($\eps^\mu = \Lambda_1^\mu$)]. The parameters $\xi=5$ and $\chi=1$ could be obtained by colliding $\unit[17]{GeV}$ photons head-on with optical ($\omega=\unit[1.55]{eV}$) laser pulses having an intensity of $\unitfrac[10^{20}]{W}{cm^2}$ (note that few-cycle laser pulses are envisaged, e.g., at the petawatt field synthesizer (PFS) in Garching \cite{ahmad_frontend_2009}). \textbf{a}) Full numerical calculation of the spectrum [see \eqref{eqn:dressedvertex_masterintegrals_wft}]. \textbf{b}) Local constant-crossed field approximation applied on the amplitude level [see \eqref{eqn:dressedvertex_linpol_semiclassicapprox_masterints}]. The inset shows that the interference pattern is lost if the local constant-crossed field approximation is applied on the probability level. \textbf{c}) Outline for $t_2=0$. Solid line: full numerical calculation; dotted (dashed) line: local constant-crossed field approximation applied on the amplitude (probability) level.}
\end{figure*}

Finally, we obtain from the stationary point $\phi_k$ [defined by \eqref{eqn:paircreation_statpointeq}] the following contribution to the master integrals [in the case of a linearly polarized laser field, see \eqref{eqn:dressedvertexmasterintegrals}] 
\begin{subequations}
\label{eqn:dressedvertex_linpol_semiclassicapprox_masterints}
\begin{gather}
\mathfrak{G}_{0}(w,t_1,t_2)
\approx
\frac{kq}{m^2} \frac{2}{w} \lsb \frac{\nfrac{w}{2}}{\abs{\chi(\phi_k)}} \rsb^{\nfrac{2}{3}} 2\pi \Ai(\rho)  \, e^{\I \widetilde{S}_\Gamma(\phi_k)},
\end{gather}
\begin{multline}
\mathfrak{G}_{1,1}(w,t_1,t_2) 
\approx
\frac{t_1}{\xi} \mathfrak{G}_{0}(w,t_1,t_2)
-
\I \lb \frac{kq}{m^2} \frac{2}{w} \rb^2  
\\\times\,
\lsb \frac{\nfrac{w}{2}}{\abs{\chi(\phi_k)}} \rsb^{\nfrac{4}{3}} \, 2\pi \Ai'(\rho) \, \psi'(\phi_k) \, e^{\I \widetilde{S}_\Gamma(\phi_k)},
\end{multline}
\end{subequations}
where $\rho = \lcb \nfrac{w}{[2\abs{\chi(\phi_k)}]} \rcb^{\nfrac{2}{3}} (1+t_2^2)$, $\chi(\phi) = \chi \psi'(\phi)$, $\chi = (\nfrac{kq}{m^2})\xi$ and $\Ai$ denotes the Airy function \cite{olver_nist_2010}. 

As the properties of the Airy function imply that pair production is exponentially suppressed for $\chi \ll 1$ and $\abs{t_2}, w\gg 1$ \cite{meuren_polarization-operator_2015,di_piazza_extremely_2012}, we will consider $\chi = 1$ in the numerical calculations. Experimentally, the regime $\chi\gtrsim 1$, $\xi\gg1$ is accessible with presently available technology, i.e.\ by colliding $\unit{GeV}$ photons (obtainable, e.g., via Compton backscattering \cite{muramatsu_development_2014,leemans_multi-gev_2014,powers_quasi-monoenergetic_2014,wang_quasi-monoenergetic_2013,kim_enhancement_2013,phuoc_all-optical_2012,esarey_physics_2009,leemans_gev_2006}) with strong optical laser pulses.

\subsection{Interference substructure of the spectrum}
\label{sec:interferencesubstructure}

To compare the results derived in \secref{sec:stationaryphaseanalysis} with those already known in the literature, we consider first a constant-crossed background field [$\psi(\phi)=\phi$]. In this case the stationary-point equation $t_1 = \xi \psi(\phi_k) = \xi \phi_k$ [see \eqref{eqn:paircreation_statpointeq}] has only one solution and the approximations leading to \eqref{eqn:dressedvertex_linpol_semiclassicapprox_masterints} are exact. Correspondingly, we obtain the probabilities (rates) for pair creation inside a constant-crossed background field given, e.g., in \cite{nikishov_pair_1967,ritus_1985} by combining \eqref{eqn:paircreation_linpoltracelambda1} with \eqref{eqn:dressedvertex_linpol_semiclassicapprox_masterints}.

For an oscillatory plane-wave background field, however, one finds in general more than one stationary point. Physically, this implies that the electron-positron pair can be created with the same asymptotic quantum numbers at different laser phases (note that different formation regions are usually separated on the macroscopic scale given by the laser wavelength). In accordance with general principles we expect that the existence of different pathways with the same final state causes interference effects similar to those in a multi-slit experiment \highlight{(for the importance of interference effects see also \cite{seipt_nonlinear_2011,mackenroth_nonlinear_2011,heinzl_beam-shape_2010,krafft_spectral_2004,narozhnyi_photon_1996,akkermans_ramsey_2012,dumlu_interference_2011,dumlu_stokes_2010,hebenstreit_momentum_2009}).} 

The presence of interference fringes in the spectrum is demonstrated in \figref{fig:spectrum_t1t2_N5interferencecompare}. The first subplot was obtained using a full three-dimensional numerical calculation of the spectrum based on the method introduced in \secref{sec:ftmasterintegrals} [see \eqref{eqn:dressedvertex_masterintegrals_wft}]. As shown in the second subplot, the semiclassical approximation introduced in \secref{sec:stationaryphaseanalysis} is already for $\xi=5$ in very good agreement with the exact result.  

Due to the fact that the background field is approximated locally around each stationary point as a constant-crossed field during the derivation of \eqref{eqn:dressedvertex_linpol_semiclassicapprox_masterints}, we call the semiclassical approximation also local constant-crossed field approximation. However, it is important that the local constant-crossed field approximation is applied on the level of the probability amplitude (i.e.\ we do not simply average the spectrum of a constant-crossed field over the laser pulse shape). The essential difference between both approaches is the presence of the phase factor $\exp{[\I \widetilde{S}_\Gamma(\phi_k)]}$ in \eqref{eqn:dressedvertex_linpol_semiclassicapprox_masterints}, which gives rise to the interference substructure once the contribution of multiple stationary points is taken into account. If the spectrum is calculated for each stationary point separately and the result is added on the probability level, the interference fringes are lost (see inset in \figref{fig:spectrum_t1t2_N5interferencecompare}; so far this approach was called local constant-crossed field approximation in the literature). 

As the interference pattern is determined by the phase factor $\exp{[\I \widetilde{S}_\Gamma(\phi_k)]}$, we conclude from \eqref{eqn:dressedvertex_reducedphase_onshell} that the oscillation frequency of the interference fringes in the spectrum scales as $\sim\xi^3$ for $w$, $\sim \xi^2$ for $t_1$ and $\sim \xi$ for $t_2$. Here, we define the oscillation frequency of the spectrum with respect to a momentum parameter $x$ as the inverse of the change $\delta x$ which is needed to advance from one local maximum of the differential probability to an adjacent one. The difference between the oscillation frequencies for $t_1$ and $t_2$ is clearly visible in \figref{fig:spectrum_t1t2_N5interferencecompare}. 

In order to fully resolve the interference substructure of the spectrum we used for $\xi\sim10$ a grid in momentum space ($w$, $t_1$, $t_2$) with $\sim 10^5 \times 10^4 \times 10^3 = 10^{12}$ data points (to obtain the two-dimensional plots we integrated numerically over the remaining momentum variable). From the above scaling laws for the oscillation frequency we conclude that this choice ensures enough sampling points per oscillation period of the interference fringes. As a cross-check we ensured that the total pair-creation probability calculated here by integrating numerically over the complete spectrum agrees with the one obtained in \cite{meuren_polarization-operator_2015} using the optical theorem. 

As the interference substructure is an intrinsic nonlocal effect, it cannot be included easily into existing PIC schemes. However, the exact resolution of the transverse momentum components is beyond the achievable precision of existing codes. Therefore, their overall precision should be increased first before interference effects can be studied.

\subsection{Classical interpretation of the stationary points}
\label{sec:classicalinterpretationofthestatpoints}

To obtain an intuitive interpretation of the stationary points discussed in \secref{sec:stationaryphaseanalysis}, we consider the classical equations of motion for an electron (positron) inside a plane-wave laser field. They predict that the time evolution of the electron four-momentum $P^\mu(\phi)$ is given by \cite{meuren_tests_2014,di_piazza_extremely_2012,meyer_covariant_1971,sarachik_classical_1970}
\begin{multline}
\label{eqn:ced_chargeinplanewavefield_momentumfourvectorsolution}
P^\mu(\phi)
=
P_0^\mu + \frac{e}{kP_0} \, \Ftilde^{\mu\nu}(\phi,\phi_0)  P_{0\nu} 
\\+
\frac{e^2}{2(kP_0)^2} \, \Ftilde^{2\mu\nu}(\phi,\phi_0) P_{0\nu},
\end{multline}
where $P_0^\mu = P^\mu(\phi_0)$ denotes the four-momentum at the laser phase $\phi_0$ and
\begingroup
\allowdisplaybreaks
\begin{multline}
\Ftilde^{\mu\nu}(\phi,\phi_0)
= 
\int^{\phi}_{\phi_0} d\phi'\, F^{\mu\nu}(\phi')
\\=
\sum_{i=1,2} f_i^{\mu\nu} [\psi_i(\phi)-\psi_i(\phi_0)]
\end{multline}%
\endgroup
the integrated field tensor [compare with \eqref{eqn:fieldtensor}]. To obtain the corresponding result for a positron we have to change the sign of the charge ($e\to-e$) in \eqref{eqn:ced_chargeinplanewavefield_momentumfourvectorsolution}. Note that a laser field does not have a dc component [$\psi_i(\pm\infty)=0$], which implies that the electron (positron) four-momentum does not change asymptotically [$P^\mu(-\infty) = P^\mu(+\infty)$]. This observation is in agreement with the Lawson-Woodward theorem \cite{lawson_lasers_1979,woodward_theoretical_1948}, which states that a plane-wave laser field cannot accelerate particles.

The classical time evolution becomes particularly transparent if the four-momentum is expanded in the canonical light-cone basis [see \eqref{eqn:canonicallcb}]
\begin{gather}
\label{eqn:classicaldynamics_momentumincanonicallcc}
P^\mu(\phi) = \rho(\phi) q^\mu + \sigma(\phi) k^\mu + m \sum_{i=1,2} \tau_i(\phi) \Lambda_{i}^\mu.
\end{gather}
The conservation of $kP=kP_0$ implies that also $\rho(\phi)=\rho(\phi_0) = \nfrac{kP}{kq}$ is conserved and 
\begin{gather}
\label{eqn:classicaldynamics_momentumincanonicallcc_onshellcondition}
\sigma(\phi) = \frac{1}{2} \frac{m^2}{kP_0} \big[1 + \tau_1^2(\phi) + \tau_2^2(\phi)\big]
\end{gather}
is determined by the on-shell condition $P^2(\phi) = m^2$. Therefore, the nontrivial dynamic is entirely described by the transverse degrees of freedom
\begin{gather}
\label{eqn:ced_planewave_electronmomentumevolutionlcc}
\tau_i(\phi) = \tau_i(\phi_0) - \xi_i [\psi_i(\phi)-\psi_i(\phi_0)].
\end{gather}
Correspondingly, also $\tau_{2}(\phi) = \tau_{2}(\phi_0)$ is conserved for a linearly polarized laser field [$\psi_2(\phi)=0$].

\begin{figure*}[t!]
\centering
\includegraphics{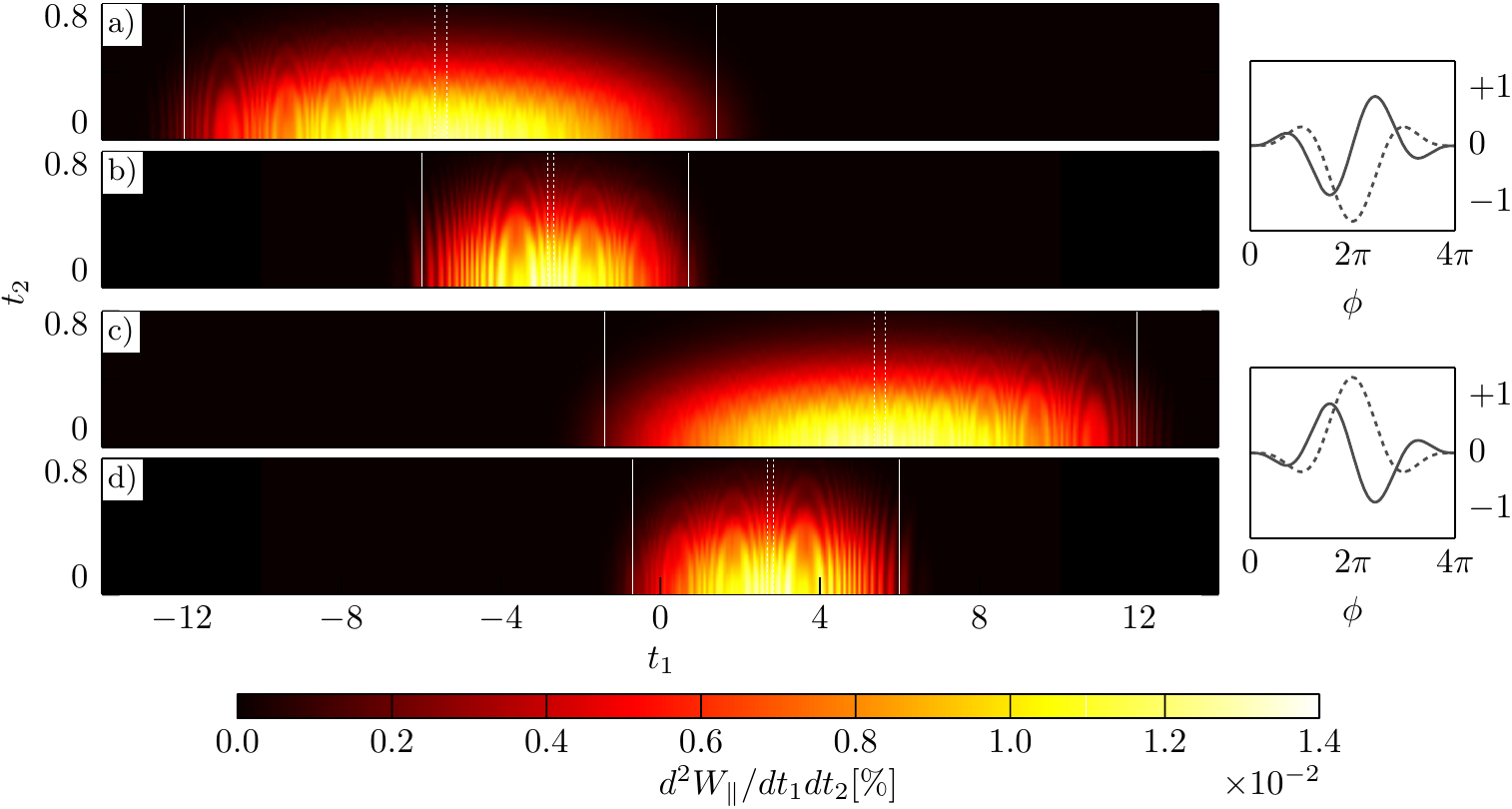}
\caption{\label{fig:spectrum_N2xicompare}\textbf{Left side}: Numerically calculated differential pair-production probability as a function of the transversal momentum parameters $t_1$ and $t_2$ (the longitudinal momentum characterized by $R$ is integrated numerically). The incoming photon has parallel polarization ($\eps^\mu = \Lambda_1^\mu$), the laser pulse $N=2$ cycles (such short pulses are envisaged, e.g., at the PFS in Garching \cite{ahmad_frontend_2009}) and $\chi=1$. We compare two different CEPs and two different intensities: \textbf{a}) $\phi_0=0$, $\xi=10$; \textbf{b}) $\phi_0=0$, $\xi=5$; \textbf{c}) $\phi_0=\pi$, $\xi=10$; \textbf{d}) $\phi_0=\pi$, $\xi=5$. The solid white lines confine the phase-space region where the pair can be produced at a phase $\phi$ with $\abs{\psi'(\phi)} \geq 0.5$ and the dashed white lines indicate the transverse momenta for which the pair can be produced at a local field peak. After integrating over $t_1$ and $t_2$ we obtain for the total pair-production probability $W_{\parallel} = 0.09\,\%$ ($\xi=10$) and $W_{\parallel} = 0.045\,\%$ ($\xi=5$), up to this precision it is independent of the CEP. \textbf{Right side}: Plot of the laser pulse shape [solid line: $\psi'(\phi)$, dashed line: $\psi(\phi)$].}
\end{figure*}

Using the above results we consider now the classical evolution of the electron and the positron four-momentum $p_1^\mu(\phi)$ and $p_2^\mu(\phi)$, respectively. Requiring the boundary conditions $p_i^\mu(\infty) = p_i$ [see \eqref{eqn:momentumincanonicallcb}], we obtain
\begin{subequations}
\label{eqn:paircreation_momentumincanonicallcb_classical}
\begin{align}
p_1^\mu(\phi) &= r' q^\mu + s'(\phi) k^\mu + m \sum_{i=1,2} t_i(\phi) \, \Lambda_{i}^\mu,  
\\
p_2^\mu(\phi) &= -r q^\mu - s(\phi) k^\mu - m \sum_{i=1,2} t_i(\phi) \, \Lambda_{i}^\mu,
\end{align}
\end{subequations}
where $r'=r+1$, 
\begin{gather}
\label{eqn:paircreation_tiphi_classical}
t_i(\phi) = t_{i} - \xi_i \psi_i(\phi)
\end{gather}
[note that $t_i=t_i(\infty)$ as $\psi_i(\infty)=0$] and [compare with \eqref{eqn:momentumincanonicallcbssprime}]
\begin{subequations}
\begin{align}
s(\phi)   &= \frac{1}{2r} \frac{m^2}{kq} [1 + t_1^2(\phi) + t_2^2(\phi)],
\\
s'(\phi)   &= \frac{1}{2r'} \frac{m^2}{kq}  [1 + t_1^2(\phi) + t_2^2(\phi)].
\end{align}
\end{subequations}
Furthermore, the following relation holds [compare with \eqref{eqn:momentumconservationpaircreation}]
\begin{gather}
\label{eqn:momentumconservationpaircreation_instantaneous}
p_1^\mu(\phi) + p_2^\mu(\phi)
=
q^\mu + Q^\mu,
\quad
Q^\mu = n(\phi) k^\mu,
\end{gather}
where
\begin{gather}
\label{eqn:momentumconservationpaircreation_instantaneous_n}
n(\phi) = \frac{w}{2} \frac{m^2}{kq} [1 + t_1^2(\phi) + t_2^2(\phi)].
\end{gather}

Assuming that a) the photon with four-momentum $q^\mu$ transforms within the short formation region $\delta\phi\sim\nfrac{1}{\xi}$ around a given laser phase $\phi_{\text{c}}$ into an electron-positron pair [see \eqref{eqn:dressedvertex_linpol_phasesemiclassicapprox} and the discussion below] and b) the charged particles subsequently obey the classical equations of motion, we conclude from \eqref{eqn:momentumconservationpaircreation_instantaneous} that the four-momentum $Q_{\text{c}}^\mu = n(\phi_{\text{c}})k^\mu$ must be absorbed ``non-classically'' from the background field during the pair-creation process itself (i.e.\ within the formation region). As the direct transformation of a real photon into a real electron-positron pair is kinematically forbidden, $n(\phi_{\text{c}})$ is always greater than zero. Stated differently, the four-momentum $Q_{\text{c}}^\mu$ is needed to bring the massive particles on shell with the right initial conditions $p_1^\mu(\phi_{\text{c}})$ and $p_2^\mu(\phi_{\text{c}})$ such that the subsequent classical propagation results in the asymptotic four-momenta $p_1^\mu=p_1^\mu(\infty)$ and $p_2^\mu=p_2^\mu(\infty)$.  

To verify the correctness of this semiclassical picture, we demonstrate now that it is in perfect agreement with the results obtained in \secref{sec:stationaryphaseanalysis} (i.e.\ it is valid in the regime $\xi\gg 1$). To this end we note the following relation [see Eqs.\sdist(\ref{eqn:paircreation_tiphi_classical}), (\ref{eqn:momentumconservationpaircreation_instantaneous_n}) and (\ref{eqn:dressedvertex_reducedphase_onshellderivative_general})] 
\begin{multline}
\label{eqn:paircreation_phaseoscillationvsn}
n(\phi) = \frac{w}{2} \frac{m^2}{kq} \Big\{ 1 +  \sum_{i=1,2} [t_i - \xi_i \psi_i(\phi)]^2 \Big\} 
\\= \widetilde{S}'_\Gamma(w,t_1,t_2;\phi).
\end{multline}
Correspondingly, the quantity $n(\phi)$, which is defined in \eqref{eqn:momentumconservationpaircreation_instantaneous_n} based on the classical equations of motion, corresponds exactly to the oscillation frequency of the master integrals [$\widetilde{S}'_\Gamma(w,t_1,t_2;\phi)$, see \eqref{eqn:dressedvertex_reducedphase_onshellderivative_general}], which are obtained from the full quantum calculation of the pair-creation probability [see \eqref{eqn:dressedvertexmasterintegrals}].

As discussed in \secref{sec:stationaryphaseanalysis}, the master integrals are dominated by small formation regions $\delta\phi\sim\nfrac{1}{\xi}$ around the (quasi-) stationary points $\phi_k$ defined by $t_{i} = \xi_i \psi_i(\phi_k)$ [i.e.\ $t_i(\phi_k) = 0$; see \eqref{eqn:paircreation_statpointeq} and \eqref{eqn:paircreation_tiphi_classical}]. This result of the stationary-phase analysis has a very intuitive explanation within the semiclassical picture developed above. According to \eqref{eqn:paircreation_phaseoscillationvsn} the stationary points correspond exactly to those laser phases where $n(\phi)$ is minimal. As the non-classical absorption of the four-momentum $Q^\mu$ can be depicted as a tunneling process (see also \cite{wollert_tunneling_2014}), it is natural to interpret $n(\phi)$ as a measure of the effective tunneling distance. Therefore, the above intuitive picture also predicts that the pair-creation process happens predominantly at those laser phases where $n(\phi)$ is minimal, in agreement with the full quantum calculation.

\subsection{Scaling laws for the spectrum}
\label{sec:spectrumscalinglaws}

From the discussion in the previous section we expect that many properties of the asymptotic momentum distribution can be understood from the classical equations of motion [see \eqref{eqn:classicaldynamics_momentumincanonicallcc}]. As an example, they predict that the spectrum extends up to $\xi_i$ in the variables $t_i$ in the general case of elliptical polarization. 

This supposition is confirmed by the stationary-phase analysis carried out in \secref{sec:stationaryphaseanalysis}. For the special case of a linearly polarized laser field [$\psi_2(\phi)=0$] we obtain $\abs{t_1} \lesssim \xi$, as for $\abs{t_1}>\xi$ the stationary-point equation $t_1 = \xi \psi(\phi_k)$ has no solution [see \eqref{eqn:paircreation_statpointeq}]. However, the differential probability with respect to $t_2$ is now entirely determined by quantum effects during the pair-creation process itself, which implies $\abs{t_2} \lesssim 1$ [$t_2$ is a constant of motion if $\psi_2(\phi)=0$]. The correctness of this bound can be seen explicitly from the argument $\rho$ of the Airy function [see \eqref{eqn:dressedvertex_linpol_semiclassicapprox_masterints} and the discussion below]. The observation that pair production is exponentially suppressed for $\abs{t_2} \gg 1$ is in agreement with the fact that the true stationary points $\varphi_k^{\pm}$ are located far away from the real axis in this case [see \eqref{eqn:paircreation_statpointslinpol}]. Both scaling laws are verified numerically in \figref{fig:spectrum_N2xicompare}.

Moreover, \figref{fig:spectrum_N2xicompare} shows that all qualitative features of the spectrum in $t_1$ except the interference substructure can be understood from the classical acceleration of the charged particles in the laser field. In particular, the position and extension of the spectrum (solid lines) is predicted well by the classical equations of motion. Furthermore, the highest pair-creation probability is obtained for those momentum parameters which require that the process happens around a peak of the laser intensity (dashed lines). 

We point out that for short laser pulses the classical acceleration has a preferred direction, which depends strongly on the CEP of the pulse. This can be seen from the plot of $\psi(\phi)$ in \figref{fig:spectrum_N2xicompare}, which determines the final momentum of the particles [see \eqref{eqn:ced_planewave_electronmomentumevolutionlcc}]. Correspondingly, the large CEP effects visible in the spectrum can be understood from classical physics (for the nonlinear Breit-Wheeler process they were first reported in \cite{krajewska_breit-wheeler_2012}).  

\begin{figure}[t!]
\centering
\includegraphics{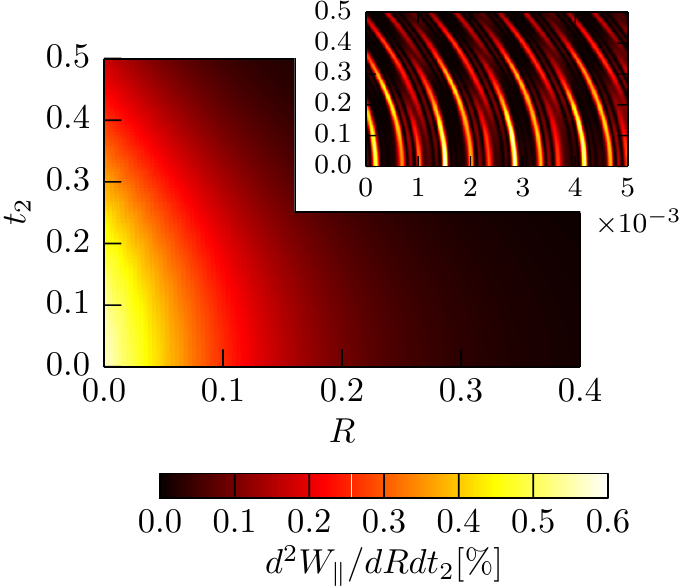}
\caption{\label{fig:spectrum_wt2}Differential probability with respect to the parameters $t_2$ and $R$ for $\chi=1$, $\xi=10$, $N=5$ and $\phi_0=\nfrac{\pi}{2}$ [full numerical calculation, $t_1$ is integrated numerically and the incoming photon has parallel polarization ($\eps^\mu = \Lambda_1^\mu$)]. The inset shows $\nfrac{d^3W_{\parallel}}{(dRdt_1dt_2)}$ for $t_1=0$ [in arb. units, see \eqref{eqn:paircreation_probabilitycanonincalparametrization}]. The pronounced interference pattern vanishes after the integral in $t_1$ is taken. Note that the (spin-summed) differential probability does not depend on the sign of $R$ and $t_2$.}
\end{figure}

For a linearly polarized background field [$\psi_2(\phi) = 0$] $R$ and $t_2$ are constants of motion [see Eqs.\sdist(\ref{eqn:momentumincanonicallcc_Rw}) and (\ref{eqn:paircreation_momentumincanonicallcb_classical})]. Therefore, the differential probability distribution with respect to $R$ and $t_2$ is entirely determined by the pair-production process itself and remains invariant under the subsequent classical propagation of the particles. This is demonstrated in \figref{fig:spectrum_wt2}. After integrating over $t_1$, the spectrum looks very similar to the one obtained for a constant-crossed field (after averaging over the laser pulse shape). In particular, we observe no substructure due to interference effects. For a fixed value of $t_1$, however, the differential spectrum shows clear interference fringes (see inset in \figref{fig:spectrum_wt2}).

Finally, we note that the scaling laws given above for the momentum variables imply that the available phase-space in $R$, $t_1$ and $t_2$ for the nonlinear Breit-Wheeler process is proportional to $\xi$. Correspondingly, the linear increase of the total pair-production probability as a function of $\xi$ in the regime $\xi\gg 1$ \cite{meuren_polarization-operator_2015} is a pure kinematic effect.

\subsection{Classical vs. quantum absorption of\\ laser four-momentum}
\label{sec:classicalvsquantumabsorption}

The results obtained in \secref{sec:classicalinterpretationofthestatpoints} [in particular \eqref{eqn:momentumconservationpaircreation_instantaneous_n}] allow us to distinguish theoretically between the four-momentum which is absorbed quantum-mechanically from the laser field during the pair-creation process itself (quantum absorption)
\begin{subequations}
\label{eqn:paircreation_nquantumvsnclassicalA}
\begin{gather}
n_{\text{q}} k^\mu = p_1^\mu(\phi_k) + p_2^\mu(\phi_k) - q^\mu
\end{gather}
and the four-momentum which is transferred classically from the laser to the charged particles during the acceleration of the particles (classical absorption) 
\begin{gather}
n_{\text{cl}} k^\mu = p_1^\mu + p_2^\mu - [p_1^\mu(\phi_k) + p_2^\mu(\phi_k)].
\end{gather}
Asymptotically, however, we observe only the sum of both processes [see \eqref{eqn:momentumconservationpaircreation}]
\begin{gather}
n = n_{\text{q}} + n_{\text{cl}}.
\end{gather}
\end{subequations}
Here, $p_1^\mu(\phi_k)$ and $p_2^\mu(\phi_k)$ [$p_{1,2}^2(\phi_k) = m^2$] denote the initial values for the classical propagation, which starts at the laser phase $\phi_k$ once the real pair is already produced [see \eqref{eqn:paircreation_momentumincanonicallcb_classical} and \secref{sec:initialconditionsclassicalpropagation}]. Correspondingly, $n_{\text{q}} k^\mu$ represents the four-momentum which must be transfered non-classically to bring the particles on shell during the creation process (see \secref{sec:classicalinterpretationofthestatpoints}) and $n_{\text{cl}} k^\mu$ denotes the four-momentum which is transfered after the production until the pair leaves the laser field with asymptotic four-momenta $p_1^\mu$ and $p_2^\mu$, respectively. Note that the classical acceleration of the particles (i.e.\ the absorption of $n_{\text{cl}}$ laser photons) does not contradict the Lawson-Woodward theorem \cite{lawson_lasers_1979,woodward_theoretical_1948}, as the charged particles are created inside the laser pulse.

In the case of linear polarization [$\psi_2(\phi)=0$] $t_2$ is a constant of motion and the stationary points are determined from $t_1 = \xi \psi(\phi_k)$ [see \eqref{eqn:paircreation_statpointeq}]. Therefore,
\begin{gather}
\label{eqn:paircreation_nquantumvsnclassical}
n_{\text{q}} = \frac{1}{2} w \frac{m^2}{kq} (1 + t_2^2),
\quad
n_{\text{cl}} = \frac{1}{2} w \frac{m^2}{kq} t_1^2.
\end{gather}
Using the scaling laws discussed in \secref{sec:spectrumscalinglaws} we conclude that the quantum and the classical absorption scale as $n_{\text{q}} \sim \nfrac{\xi}{\chi}$ and $n_{\text{cl}} \sim \nfrac{\xi^3}{\chi}$, respectively. 

An intuitive understanding of the scaling law for $n_{\text{q}}$ is obtained by squaring \eqref{eqn:paircreation_nquantumvsnclassicalA}, which shows that
\begin{gather}
n_{\text{q}} \geq 2\frac{m^2}{kq}.
\end{gather}
This lower bound becomes an equality at the pair-production threshold, which is obtained for $p_1^\mu(\phi_k)= p_2^\mu(\phi_k)$ (i.e.\ for $w=4$ and $t_2=0$ in the case of linear polarization).

Note that although we indicated $n_{\text{q}}$ as the number of ``quantum-⁠absorbed'' laser photons (referring to the fact that the electron and the positron are virtual inside the formation region), the scaling law $n_{\text{q}} \sim \xi$ (for $\chi\gtrsim 1$) can be understood by noting that the four-momentum which is transferable classically during the formation region $\delta\phi \sim \nfrac{1}{\xi}$ [see \eqref{eqn:dressedvertex_linpol_phasesemiclassicapprox} and the discussion below] scales as $\sim \xi k^\mu$ [see \eqref{eqn:classicaldynamics_momentumincanonicallcc} and \eqref{eqn:momentumconservationpaircreation_instantaneous_n}].

In conclusion, the classical energy transfer during the propagation of the particles is much larger than the energy transfer which takes place during the pair-creation process itself in the regime $\xi\gg 1$ ($n_{\text{cl}} \gg n_{\text{q}}$). Correspondingly, a possible depletion of the background field (e.g., due to the development of a QED cascade) is dominated by the classical energy transfer from the laser to the pairs.

\subsection{Initial conditions for the classical propagation}
\label{sec:initialconditionsclassicalpropagation}

The semiclassical picture established in \secref{sec:classicalinterpretationofthestatpoints} agrees with the basic principles used in PIC codes. In order to include the Breit-Wheeler process into a PIC code, the main difficulty is to determine the correct initial conditions for the classical propagation of the charged particles. To this end the spectrum obtained for a constant-crossed field is usually employed (see \secref{sec:stationaryphaseanalysis}). As $R$ and $t_2$ are constants of motion [see Eqs.\sdist(\ref{eqn:momentumincanonicallcc_Rw}) and (\ref{eqn:paircreation_momentumincanonicallcb_classical})], the asymptotic momentum distribution in $R$ and $t_2$ obtained for a constant-crossed field using the $S$-matrix approach agrees with the initial momentum distribution in $R$ and $t_2$ immediately after the particles are brought on shell. Therefore, this approach is reasonable for the two parameters $R$ and $t_2$.

In contrast, the appropriate initial condition for $t_1(\phi_k)$ is far from clear, as $t_1(\phi_k)$ is not conserved by the classical equations of motion. At first sight one could expect the existence of a distribution function which determines the initial condition for $t_1$ (similar as for $R$ and $t_2$). However, we showed in \secref{sec:classicalinterpretationofthestatpoints} that the fixed value $t_1(\phi_k)=0$ should be used. This condition defines the stationary points [for linear polarization, see \eqref{eqn:paircreation_statpointeq}] and therefore leads to \eqref{eqn:dressedvertex_linpol_semiclassicapprox_masterints} which represent the probabilities inside a locally constant-crossed field. Correspondingly, the usage of \eqref{eqn:dressedvertex_linpol_semiclassicapprox_masterints} to determine the probability distributions for $R$ and $t_2$ requires the initial condition $t_1(\phi_k)=0$ for the classical propagation.

\section{Summary and conclusions}

In the present paper the momentum distribution for electron-positron pairs produced via the nonlinear Breit-Wheeler process has been investigated for short laser pulses. Using a newly developed numerical scheme \highlight{(see \secref{sec:ftmasterintegrals})} we have calculated for the first time the spectrum on a three-dimensional lattice which fully resolves the interference substructure even in the ultra-relativistic regime $\xi \gg 1$. Furthermore, we have investigated the local constant-crossed field approximation and showed that it reproduces the spectrum (including the interference fringes) for $\xi \gg 1$ if it is applied on the probability-amplitude level. Correspondingly, three effects determine the final momentum distribution in the regime $\xi\gg 1$: The production of an electron and a positron with physical mass $m$ inside a constant-crossed field, their subsequent classical acceleration by the laser field and the interference between all production channels which lead to the same asymptotic quantum numbers. Accordingly, we verified that the produced electron and positron behave like classical particles after they have left the formation region \highlight{(see \secref{sec:semiclassicalpicture}, in particular \secref{sec:classicalinterpretationofthestatpoints} and \secref{sec:spectrumscalinglaws})} and that the substructure of the spectrum can be attributed to interferences between the contributions of different formation regions similar to those in a multi-slit experiment \highlight{(see \secref{sec:interferencesubstructure})}. Furthermore, it is shown that one can distinguish between a classical and a quantum absorption of laser photons \highlight{(see in particular \secref{sec:classicalvsquantumabsorption})}. As the former is dominant in the regime $\xi \gg 1$, a possible depletion of the laser field during the development of a QED cascade is mainly caused by the classical acceleration of the created charged particles. \highlight{In summary, the new findings presented here allow a disentanglement between classical and quantum aspects of the pair-production process.}

\section*{Acknowledgement}

S.M. would like to thank Alexander M. Fedotov, Thomas Grismayer, Karen Z. Hatsagortsyan, Oleg Skoromnik, Matteo Tamburini, Marija Vranic, Anton W\"ollert and Enderalp Yakaboylu for fruitful discussions. Furthermore, he is grateful to the Studienstiftung des deutschen Volkes for financial support. All plots have been created with Matplotlib \cite{hunter_matplotlib_2007} and the GSL \cite{GSL} has been used for numerical calculations.

\textit{Note\,added.---} After the publication of the first preprint of this paper the references \cite{seipt_caustic_2016,nousch_spectral_2016,seipt_analytical_2016,dinu_quantum_2016,jansen_strong-field_2015} appeared, which cover related aspects. The authors are very thankful to Daniel Seipt for insightful discussions about the findings reported in Refs. \cite{seipt_caustic_2016,nousch_spectral_2016}.

\appendix

\section{Light-cone coordinates}
\label{sec:lccappendix}

We call the four four-vectors $k^\mu$, $\bar{k}^\mu$, $e_i^\mu$ ($i\in 1,2$) a light-cone basis if they obey \cite{meuren_polarization_2013}
\begin{gather}
\label{eqn:lc_basisproperties}
\begin{gathered}
k^2 = \bar{k}^2 = 0, 
\quad
ke_i = \bar{k}e_i = 0,
\\
k\bar{k} = 1, 
\quad
e_i e_j = - \delta_{ij}.
\end{gathered}
\end{gather}
Using the above properties and the determinant identity for $\eps^{\mu\nu\rho\sigma}\eps^{\alpha\beta\gamma\delta}$ one finds that any such light-cone basis obeys $\Omega^2 = 1$, where
\begin{gather}
\label{eqn:lcc_orientation}
\Omega = \eps_{\mu\nu\rho\sigma} k^\mu \bar{k}^\nu e_1^\rho e_2^\sigma
\end{gather}
is called the orientation of the basis. In light-cone coordinates the metric $g^{\mu\nu}=\diag(+1,-1,-1,-1)$ is given by
\begin{gather}
g^{\mu\nu} = k^\mu \bar{k}^\nu + \bar{k}^\mu k^\nu - e_1^\mu e_1^\nu - e_2^\mu e_2^\nu.
\end{gather}

Intrinsically, the photon decay inside a plane-wave field [see \secref{sec:ppp}, in particular \eqref{eqn:fieldtensor}] is characterized by the two light-like four-vectors $k^\mu$ and $q^\mu$ and the constant field tensors $f_i^{\mu\nu}$. Therefore, it is natural to expand the four-momenta of the created electron-positron pair in the following light-cone basis
\begin{gather}
\label{eqn:canonicallcb}
k^\mu,
\quad
q^\mu,
\quad
\Lambda_{1}^\mu = \frac{f^{\mu\nu}_1 q_{\nu}}{kq\, \sqrt{-a_1^2}},
\quad
\Lambda_{2}^\mu = \frac{f^{\mu\nu}_2 q_{\nu}}{kq\, \sqrt{-a_2^2}}
\end{gather}
(we will always excluded the trivial case $kq=0$ where pair production is forbidden and assume that $kq\neq 0$), which fulfills the completeness relation 
\begin{gather}
\label{eqn:canonicallcb_completenessrelation}
g^{\mu\nu}= \frac{1}{kq} (k^{\mu}q^{\nu} + q^{\mu}k^{\nu})-\Lambda_1^\mu\Lambda_1^\nu-\Lambda_2^\mu\Lambda_2^\nu
\end{gather}
\highlight{(the two four-vectors $\Lambda_{1,2}^\mu$ have been used previously by several authors to analyze processes within strong plane-wave background fields, see, e.g., \cite{baier_interaction_1975,becker_vacuum_1975})}. To distinguish the set of four four-vectors given in \eqref{eqn:canonicallcb} from other light-cone bases we will call it the canonical light-cone basis.

As any set of four four-vectors $e_i^\mu$ ($i\in 1,2$), $\bar{k}^\mu$ and $k^\mu$ which obeys the relations given in Eq. (\ref{eqn:lc_basisproperties}) represents a light-cone basis, it is natural to ask which expressions are invariant under a change of the underlying light-cone basis. To this end we consider two different bases $\bar{k}^\mu$, $e_i^\mu$ and $\bar{k}'^\mu$, $e_i'^\mu$ and denote the corresponding components of a four-vector $v^\mu$ by 
\begin{gather}
\begin{aligned}
v^\lplus &= \bar{k}^\mu v_\mu,&
v^\lone  &= e_1^\mu v_\mu,&
v^\ltwo  &= e_2^\mu v_\mu,&
\\
v'^\lplus &= \bar{k}'^\mu v_\mu,&
v'^\lone  &= e_1'^\mu v_\mu,&
v'^\ltwo  &= e_2'^\mu v_\mu,&
\end{aligned}
\end{gather}
($v^\lminus = v'^\lminus = kv$). The three coordinates $(\lminus,\lperp = \lone,\ltwo)$ define a closed subspace and we obtain the relation
\begin{gather}
\begin{pmatrix}
v'^\lminus \\ v'^\lone \\ v'^\ltwo
\end{pmatrix}
=
\begin{pmatrix}
1 & 0 & 0 \\
e_1'\bar{k} & -e_1'e_1  & -e_1'e_2 \\
e_2'\bar{k} & -e_2'e_1 & -e_2'e_2  \\
\end{pmatrix}
\cdot
\begin{pmatrix}
v^\lminus \\ v^\lone \\ v^\ltwo
\end{pmatrix}.
\end{gather}
In order to show that its determinant has magnitude one, we write
\begin{gather}
\begin{aligned}
e_1'^\mu &= a e_1^\mu + b e_2^\mu + \lambda k^\mu,\\
e_2'^\mu &= c e_1^\mu + d e_2^\mu + \mu k^\mu.
\end{aligned}
\end{gather}
As $e_1'^2 = e_2'^2 = -1$ and $e_1'e_2'=0$, we obtain $a^2+b^2=c^2+d^2=1$ and $ac + bd = 0$. Without restricting generality, we set $a=\cos\varphi$, $b=\sin\varphi$ and $d=\cos\theta$, $c=\sin\theta$. Finally, we obtain the two solutions $\theta = - \varphi$ and $\theta = - \varphi+\pi$, which correspond to $ad - bc = \pm 1$. Therefore, the measure $dv^\lminus dv^\lperp=dv'^\lminus dv'^\lperp$ and the delta function 
\begin{multline}
\label{eqn:lcc_momemtumdeltafunctionsnotation}
\delta^{(\lminus,\lperp)}(v)
=
\delta(v^\lminus) \delta(v^\lone) \delta(v^\ltwo)
\\=
\delta(v'^\lminus) \delta(v'^\lone) \delta(v'^\ltwo)
=
\delta^{(\lminus,\lperp)}(v')
\end{multline}
are invariant under a change of the light-cone basis. If $v^\lminus=v^\lperp=0$ also the component $v^\lplus = \bar{k}^\mu v_\mu$ is invariant.

\section{Photon polarization density matrix}
\label{sec:photonpolarizationappendix}

The (complex) polarization four-vector $\eps^\mu$ of a photon with four-momentum $q^\mu$ ($q^2=0$) must obey $\eps^{*\mu} \eps_{\mu} = -1$ and $q\eps=0$. In the light-cone basis $k^\mu$, $q^\mu$, $\Lambda^\mu_i$ [see \eqref{eqn:canonicallcb}] we obtain for the metric [see \eqref{eqn:canonicallcb_completenessrelation}]
\begin{gather}
g^{\mu\nu} = \frac{1}{kq} \lb k^\mu q^\nu + q^\mu k^\nu \rb - \Lambda_1^\mu \Lambda_1^\nu - \Lambda_2^\mu \Lambda_2^\nu
\end{gather}
and thus the polarization four-vector is given by
\begin{gather}
\begin{gathered}
\eps^\mu = c_1 \Lambda_1^\mu + c_2 \Lambda_2^\mu + c_3 q^\mu,
\\
c_1 = - (\eps\Lambda_1),
\quad
c_2 = - (\eps\Lambda_2),
\quad
c_3 = \frac{k\eps}{kq}
\end{gathered}
\end{gather}
with the normalization condition $\abs{c_1}^2 + \abs{c_2}^2 = 1$. As the contraction of the matrix element with the four-momentum $q^\mu$ must vanish due to gauge symmetry, we can restrict us to the vectors $\Lambda_i^\mu$ and replace the density matrix by
\begin{subequations}
\begin{gather}
\rho^{\mu\nu} = \eps^{\mu} \eps^{*\nu} 
\longrightarrow
\sum_{i,j=1,2} \rho_{ij} \Lambda_i^\mu \Lambda_j^\nu,
\\
\begin{gathered}
\rho_{11} = \abs{c_1}^2,
\quad
\rho_{22} = \abs{c_2}^2,
\\
\rho_{12} = c_1c_2^*,
\quad
\rho_{21} = c_1^*c_2.
\end{gathered}
\end{gather} 
\end{subequations}
The $2\times 2$ density matrix $\rho_{ij}$ is  Hermitian and has unit trace
\begin{gather}
\begin{gathered}
\rho_{ij}
=
\Lambda_{i\mu}\Lambda_{j\nu} \rho^{\mu\nu},
\quad
\rho^\dagger_{ij} = \rho_{ji}^* = \rho_{ij},
\\
\tr \rho = \sum_{i=1,2} \rho_{ii} = 1.
\end{gathered}
\end{gather}
Any Hermitian $2\times 2$ matrix can be expanded using the Pauli matrices $\sigma^i$ and the identity $\one$ (with real parameters). Since $\tr\sigma^i = 0$, we obtain (see \cite{landau_quantum_1981}, Eq. 8.9)
\begin{multline}
\rho 
= 
\frac{1}{2} (\one + s_i \sigma^i) = \frac{1}{2} \begin{pmatrix} 1 + s_3 & s_1-\I s_2\\ s_1+\I s_2 & 1-s_3\end{pmatrix}
\\=
\begin{pmatrix} \begin{aligned}& \abs{c_1}^2 & c_1c_2^* & \\ & c_1^*c_2 & \abs{c_2}^2 & \end{aligned}\end{pmatrix},
\end{multline}
where $s_i$ are called Stokes parameters. The Stokes vector $\spvec{s} = (s_1,s_2,s_3)$ is a unit vector, which can be seen from
\begin{gather}
\det \rho = 0 = \frac{1}{4} \lb 1 - \spvec{s}^2 \rb.
\end{gather}
Correspondingly, it can be described by two Stokes angles 
\begin{gather}
\begin{gathered}
s_1 = \cos(\varphi)\sin(\theta),
\quad
s_2 = \sin(\varphi)\sin(\theta),
\\
s_3 = \cos(\theta).
\end{gathered}
\end{gather}
Using the trigonometric identities
\begin{gather}
\begin{gathered}
\frac{1}{2} (1 + \cos\theta) 
=
\cos^2(\nfrac{\theta}{2}),
\\
\frac{1}{2} (1 - \cos\theta) 
=
\sin^2(\nfrac{\theta}{2}),
\\
2\cos(\nfrac{\theta}{2}) \sin(\nfrac{\theta}{2}) = \sin (\theta)
\end{gathered}
\end{gather}
we conclude that the complex coefficients $c_1$ and $c_2$ can be expressed in terms of the stokes angles as
\begin{gather}
\label{eqn:stokesparamters_abvsphitheta}
c_1 = \cos(\nfrac{\theta}{2}) \, e^{-\I \nfrac{\varphi}{2}},
\quad
c_2 = \sin(\nfrac{\theta}{2}) \, e^{+\I \nfrac{\varphi}{2}},
\end{gather}
implying the representations
\begin{gather}
\label{eqn:stokesparameters_abvsxi}
\begin{gathered}
\abs{c_1}^2 = \cos^2(\nfrac{\theta}{2}),
\quad
\abs{c_2}^2 = \sin^2(\nfrac{\theta}{2}),
\\
c_1c_2^* = \frac{1}{2} \sin(\theta) \, [\cos(\varphi) - \I \sin(\varphi)]
\end{gathered}
\end{gather}
[note that we can always multiply by a total phase in Eq. (\ref{eqn:stokesparamters_abvsphitheta})].

\end{document}